\documentclass{aa}  
\usepackage{graphicx}
\usepackage{natbib}
\bibpunct{(}{)}{;}{a}{}{,}
\usepackage{aalongtable}
\usepackage{txfonts}
\begin{document}
   \title{On the origin of field O-type stars}

\author{E.~Schilbach  \and S.~R\"{o}ser   }

   \offprints{S.~R\"{o}ser}

\institute{Astronomisches Rechen-Institut, Zentrum f\"ur Astronomie der Universit\"at Heidelberg,  M\"{o}nchhofstra\ss{}e 12-14,
D--69120 Heidelberg, Germany\\
email: elena@ari.uni-heidelberg.de, roeser@ari.uni-heidelberg.de}
   \date{Received April 9, 2008; accepted May 30, 2008}

 
\abstract
   {}   
   {We try to identify the origins  of field O-stars in the nearest 
   2 to 3~kpc around the Sun using the best presently available kinematic data on 
   O-stars and on young open clusters. We investigate the question if the present-day data
   are consistent with the assumption that O-stars
   have formed in groups (clusters, associations),  or in isolation.}
   {We apply the epicycle theory for back-tracing the orbits of O-type stars
   and of candidate parent open clusters.}
   {From the 370 O-stars in the ``Galactic O star catalog v 2.0'' (GOSV2) we have 
   investigated 93 stars classified as  $field$, and found the origin for 73 
   of them in 48 open clusters younger than 30 Myrs. Only for 32 stars or 
   about 9\% of all O-stars from this catalogue, the question of their origin
   in groups is not solved; some of them may have originated in isolation or 
   may have disintegrated the group in which they formed.
   Fifty percent of the young  open clusters (age $<$ 30 Myr)
   in the ``Catalogue of Open Cluster Data'' (COCD) have O-stars 
   as members, or have ejected at least one O-star in the first 10 Myrs of 
   their life, or both. During this period the average mass loss from open 
   clusters by ejecting O-stars is found to be  3 to 5~$M_\odot$ per Myr.
   We prove that $\zeta$ Pup had its origin in the open cluster Trumpler~10  which 
   it left about 2.5 Myrs ago, and that its present-day distance is
   300 pc (compared to 440~pc before). The revised distance implies a 
   significant revision of the stellar parameters (a radius of 14~$R_\odot$, 
   a mass of 22.5~$M_\odot$, and a luminosity of log $L/L_\odot$ of 5.74) i.e,
   $\zeta$ Pup is closer, less massive, and less luminous than previously thought.
   Our findings provide independent estimates of the present-day distances 
   and absolute magnitudes of field O-stars.} 
   {}
   \keywords{Stars: early-type -- Stars: formation -- open clusters and associations:
general}    
   \maketitle
%

\section{Introduction}~\label{intro}

Do all O-stars form in groups (clusters, associations) as is commonly believed
or is the formation
of O-stars in isolation possible? This long-standing question can only be answered,
when the birth-places of all O-stars will be discovered. A review of the
situation is given in the introduction by  \citet{1987ApJS...64..545G} and recently
in \citet{2007ARA&A..45..481Z}. \citet{1987ApJS...64..545G} compilied a catalogue
of 195 O-stars which he used to infer the first solid statistics about runaway and field
O-stars. Recently, a new catalogue of Galactic O-stars  
(GOSV1 version 1, \citet{2004ApJS..151..103M}; GOSV2 version 2, \citet{2007astro.ph..3005S}) was published.
Comprising 370 O-stars, the catalogue allows to re-address the statistics of O-star
birth-places. In particular, the GOSV2 catalogue contains a subset of 105 O-stars 
called {\em field}, which simply means that they cannot be identified as present or former members
of recognised groups. Such ''isolated O-stars'' are of essential interest to decide the question
if ''isolated'' massive star formation is possible or not. 

Because of the relatively short lifetime (a few million years) near the main sequence, the
orbit of an O-type star in the wider solar neighbourhood can, in principle, be 
followed all the way back in time
to the onset of its hydrogen-burning stage. This means that the location of its parent star
forming cloud can be determined. Investigating the area around these parent clouds, 
one may find other young objects there, e.g. young star clusters or OB-associations.

During the last decade, after the results of the Hipparcos mission became available,
nearby OB-associations have been investigated in considerable detail 
\citep[][]{1999AJ....117..354D}.
However, not all OB-stars have been found living in associations,
some are far way from presently-known stellar groups on the sky.

Using the data from ESA's Hipparcos mission, \citet{2001A&A...365...49H} back-traced the orbits
of 56  OB-type runaway stars and nine compact objects with distances less than 700 pc.
They found that at least 21 objects of their sample could be linked back to nearby
associations and young open clusters. The authors state that the remaining objects
may have originated from distances farther away than 700 pc, where the knowledge of
parent groups is poor.

Another line of argument has been followed in two papers by \citet{2004A&A...425..937D,2005A&A...437..247D}.
In their first paper they investigate the origin of 43 O-type {\em field} stars from
the O-star catalogue by \citet{1987ApJS...64..545G}. The authors search the area around
these stars for stellar groups in the near-infrared which are possibly hidden in the optical.
In their second paper  \citep{2005A&A...437..247D} they investigate the same sample kinematically.
They excluded as {\em field} stars the runaway stars, i.e. those with spatial velocities
above the limit of 40~km~s$^{-1}$ set by \citet{1961BAN....15..265B} and those at distances
larger than 250~pc from the Galactic
plane. Combining the results of both papers
they claim that not more than 4 $\pm$ 2\% of all O-stars in Gies'
catalogue can be called genuine {\em field} stars.

The argument by \citet{2004A&A...425..937D,2005A&A...437..247D} is a rather indirect 
one, they are excluding stars as  {\em field}
without being able to retrace their origin. This is exactly the point where we start
our present study. Only if one succeeds to retrace an O-star to a parent group within its
past lifetime one can say with certainty that this O-star has originated in a group.
Proving or disproving this point is not an easy task given our incomplete knowledge of possible
birth-places in the wider neighbourhood of the Sun and the uncertainties of the
six-dimensional phase
space coordinates (position and motion) of candidate stars and candidate clusters and/or
associations.

In this paper we are testing the hypothesis that O-stars, the origin of which is hitherto
unknown, may have been ejected from young open clusters (or their protoclusters)
during or after the star formation
period in the (parental) cluster. For this purpose, we follow the path of stars and clusters
back in time in the Galactic potential. In the next section we present the underlying observations,
then we describe the method and its  application. Section~\ref{res} is a presentation
and a discussion of the results. In Section~\ref{indstars} we consider a few selected examples
of stars with the adopted solutions, whereas in Section~\ref{nosolu} we briefly discuss 
the stars for which we did not find an acceptable solution.
A summary concludes the paper.

\section{Observational material}~\label{obs}

For the back-tracing of stellar and cluster orbits we are using the most homogeneous
and accurate
data of all 6 parameters of space phase available at present.
We took the positions and proper motions from the recently completed 
PPMX catalogue \citep{PPMX}, and the radial velocities
from the CRVAD-2 \citep{2007AN....328..889K}. The major sources of
specific information on open clusters and O-type stars  were the catalogues
by \citet{2005A&A...438.1163K,2005A&A...440..403K} and \cite{2007astro.ph..3005S}, respectively.

\subsection{Open clusters}~\label{obs_oc}

The Catalogue of Open Cluster Data (COCD) and its Extension \citep{2005A&A...438.1163K,2005A&A...440..403K}
includes 641 open clusters and 9 cluster-like associations identified in the
ASCC-2.5 catalogue \citep{kha01}. 
For each cluster the membership
was determined using spatial, kinematic, and photometric criteria 
\citep{2004AN....325..740K}.
A homogeneous set of cluster parameters was 
derived by applying a uniform technique. The nine associations are included
in the COCD because of their compact appearance on the sky and
small dispersion in proper motion space, so that they can be treated via our
standard membership selection procedure. The COCD contains the celestial position of
a cluster, its distance to the Sun, reddening, age, angular size, proper
motions, and, if available, radial velocity. Recently, the parameter set was
supplemented by tidal radii and masses \citep{2008A&A...477..165P}. The
completeness of the cluster sample is mainly defined by the limiting magnitude
of $V \approx 11.5$ of the ASCC-2.5. Therefore, even nearby  embedded clusters 
can been missing in the sample if their members were fainter than 
$V \approx 11.5$ in the optical. Nevertheless, for ``classical''
open clusters, i.e. when the bulk of the placental matter is removed and clusters
become  visible in the optical spectral range, the sample was found to
be complete up to a distance of about 850~pc \citep{2006A&A...445..545P}.

\begin{figure}
\resizebox{\hsize}{!}{
\includegraphics[bb=90 48 544 680,width=9cm,clip,angle=270]{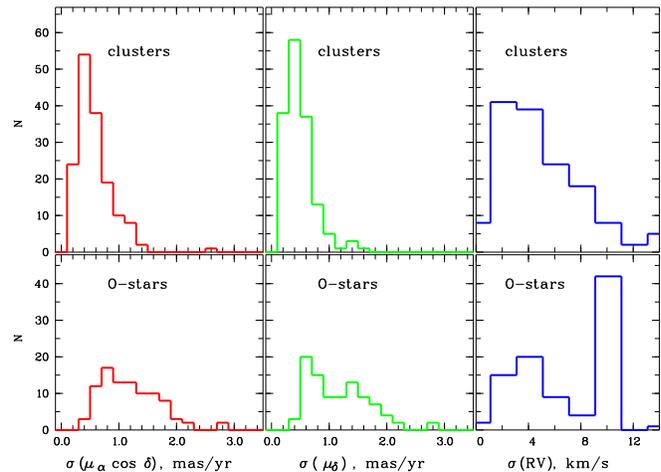}}
\caption{The distribution of the $rms$ errors of proper motions $\mu_{\alpha}\cos \delta$ (left),
$\mu_{\delta}$ (middle), and of radial velocities $RV$ (right). Upper panels show
the distributions for the 161 young open clusters, lower panels for the 93 field O-stars. The peak at 10 km/sec
in $\sigma_{RV}$ of O-stars represents stars for which no information on the $rms$ error
of the radial velocity is given in the literature.
}\label{fig:kin_cs}
\end{figure}

In the context of this paper, however, we are not interested in the full sample, 
but in a sub-sample of young clusters. As potential candidates for parent groups
we considered 161 clusters (including 9  associations) younger than 30~Myr,
and having radial velocities measured. Since absolutely bright stars are still
present in these clusters, this subset is volume limited to about
2~kpc \citep{2006A&A...445..545P}, or to a distance modulus $(V-M_V) \approx 13$
\citep{2006A&A...456..523S} when extinction is taken into account. Except for proper motions, the cluster data for each
cluster (i.e., coordinates of the cluster centre, distance, and radial velocity) were
taken from the COCD. The mean proper motions were recomputed
from the PPMX data and the membership information given in \citet{2004AN....325..740K}.
In the upper panels of Fig.~\ref{fig:kin_cs},  we show histograms of the mean errors of
the kinematic data for this sub-sample of young clusters.

\subsection{O-type stars}~\label{obs_os}

The sample of O-type stars for this paper was taken from the second
version GOSV2 \citep[][]{2007astro.ph..3005S} of the  
``Galactic O star catalog'' 
by \cite{2004ApJS..151..103M}, with 370 entries.
The catalogue is expected to be complete for O-stars brighter than $V = 8$ but
it includes many fainter stars, too. For each star, the catalogue delivers
spectral classification, photometric and astrometric data, and further
information such as multiplicity and membership in known associations.
Among 370 stars, 105 stars are classified as field stars or field runaways 
from unknown parent groups. These are the stars we were interested in.

Again, we took the positions and proper motions from the PPMX catalogue.
Doing so, we not only benefitted from accurate proper motions but also kept
proper motions of clusters and O-stars on the same system. Radial velocities
came from CRVAD-2. Though CRVAD-2 presents data for 55000 stars, 
radial velocities are available for only 93 O-stars of our list. Unfortunately, for 41 stars,
no information on the $rms$ errors of radial velocities is available, so we assumed them to
be $\pm 10$~km/s. The histograms of the mean errors
of kinematic data for the final sample of O-stars are shown
in the lower panels of Fig.~\ref{fig:kin_cs}. As expected, the accuracy of the
data for clusters is, on average, higher than for O-stars. For clusters, the 
median of the $rms$ errors of $\mu_{\alpha}\cos \delta$, $\mu_{\delta}$ and $RV$
are 0.5 mas/yr, 0.4 mas/yr and 3.4 km/s, respectively. For O-stars, the
corresponding numbers are 1.2 mas/yr, 1.0 mas/yr and 7.5 km/s.

The distances of O-stars are more of a problem. The GOSV2 
gives Hipparcos \citep{1997yCat.1239....0E} parallaxes for all stars for which these are available.
Very recently, \citet{2007hnrr.book.....V} published the new reduction of the 
Hipparcos observations. Due to a
sophisticated modelling of the satellite's attitude, \citet{2007hnrr.book.....V}
could considerably improve the trigonometric parallaxes of stars brighter 
than about 8th visual magnitude. However, only five O-stars of our sample had Hipparcos parallaxes
with an accuracy better than 30\%.

In order to get distances for all stars, there was no other chance than using
the methods of distance estimates based on the spectroscopic and photometric data.
To derive spectroscopic distances $d_{sp}$ from the well known relation
\begin{equation}
\log d_{sp} = 0.2[V - M_V +5 -3.1((B-V) - (B-V)_0)], 
\end{equation}
we took the observed $B$ and $V$ magnitudes and
the spectral classification from the GOSV2 catalogue, and converted spectral type into
absolute magnitude M$_V$ and (B-V)$_0$ according to \citet{schmidtkaler}.
However, the spectral classification of O-stars is neither straightforward nor unambiguous. For example,
for a relatively bright star, HD 135240 ($V$ = 5.08), one finds a spectral type and
luminosity class
of O~7.5\,III in the GOSV2 catalogue and O~8.5\,V in SIMBAD. This difference in 
spectral classification leads to an uncertainty of about 1~mag in distance modulus.
The second source of uncertainty arises from the $M_V$--spectral-type calibration
which can introduce systematic effects of up to 1~mag \citep[e.g., see][]{2002AJ....124..507W}. Moreover,
the calibration of O-stars shows a large scatter which may be intrinsic to the
stars themselves \citep[e.g.,][]{1983ApJ...274..302C,1992A&AS...94..211G}. 
We conclude that the spectroscopic distance moduli may be uncertain by up
to 2~mag, and the distance of a field O-star is the most inaccurate input parameter
in the back-tracing of stellar and cluster orbits.

\section{The back-tracing method}~\label{meth}

\subsection{The epicycle approach}~\label{epic}

For the re-tracing of the stars and open clusters we followed the approach
used by \cite{2006MNRAS.373..993F} who adopted the epicyclic equations of
motion as given by \citet{1959HDP....53...21L}:
\begin{eqnarray}
\xi(t)  & = &\xi(0) + \frac{\upsilon (0)}{2B}[1-\cos (\kappa t)] +
\frac{u(0)}{\sin (\kappa t)},\nonumber\\
\eta(t) & = &\eta(0) + 2A\,[\xi(0) + \frac{\upsilon (0)}{2B}]\, t -
\nonumber\\
        &   & -  \frac{\Omega_0}{B \kappa}\upsilon  (0) \sin (\kappa t) +
\frac{2\Omega_0}{\kappa2} u(0)[1-\cos (\kappa t)], \\
\zeta(t)& = &\zeta(0) \cos(\nu t) + \frac{w(0)}{\nu} \sin(\nu t),\nonumber
\end{eqnarray}

\begin{eqnarray}
u(t)    & = &u(0)\cos (\kappa t) + \frac{\kappa}{2B}\upsilon (0)\sin
(\kappa t), \nonumber\\
\upsilon (t)    & = &-\frac{2B}{\kappa}u(0)\sin (\kappa t) + \upsilon
(0)\cos (\kappa t), \\
w(t)    & = &w(0)\cos(\nu t) - \zeta(0)\nu \sin(\nu t)\nonumber
\end{eqnarray}
where $A$ and $B$ are Oort's constants, $\Omega_0$ is the angular velocity
of the Galactic rotation of the local standard of rest (LSR),  $\kappa$ and $\nu$
are the epicycle frequency and the vertical oscillation frequency,
respectively.  We assumed a flat rotation curve $A = -B = \Omega_0/2$, 
where $\Omega_0 = 25.9$~km/s/kpc at the Galactocentric radius of the LSR
$r_0 = 8.5$~kpc. We
also adopted the same values as \cite{2006MNRAS.373..993F} for the
parameters $\kappa$ = 39.0 km/s/kpc and $\nu$ = 74 km/s/kpc.

Eqs. (2) and (3) describe the motion of a particle in a non-inertial
coordinate system centred at a fiducial point at a Galactocentric
radius $r_0$ (at Z=0) from the Galactic centre, for which the
transformation from the Cartesian Galactic coordinates $X$, $Y$, $Z$
into $\xi, \eta, \zeta$ is given by
\begin{eqnarray}
\xi  & = & r_0 - r, \nonumber\\
\eta & = & r_0 \times \arctan(Y/r), \\
\zeta  & = & Z ,\nonumber
\end{eqnarray}
with $ r = \sqrt{(r_0-X)^2 + Y^2}$. Similarly, the velocity components
$U$, $V$, $W$ of the peculiar space velocity (after correcting for
solar motion and Galactic rotation) are transformed into $u$, $\upsilon$,
$w$ via
\begin{eqnarray}
u & = & U \frac{r_0-X}{r} - V \frac{Y}{r}, \nonumber\\
\upsilon & = & U \frac{Y}{r} + V \frac{r_0-X}{r}, \\
w & = & W .\nonumber
\end{eqnarray}

As we are only interested in the relative location and velocity of a star
with respect to a candidate parent cluster, we 
chose $r_0$ close to the Galactocentric radius of the candidate cluster in 
each case, and the flat rotation curve gave $\Omega_0 = 220/r_0$~km/s/kpc. 
The epicycle and the oscillation frequencies have been assumed to be constant. 
This extends the validity of the approach,
as the initial $\xi, \eta$ are small for star and cluster, even at larger
distances from the Sun, provided that $r_{star} - r_0 << r_0$, and
$ \upsilon << r_0\,\Omega_0$. We checked that the requirements were fulfilled for
our solutions.

\subsection{Variation of the initial conditions and selection of the solution}~\label{proc}

As a starting point of the backward computations, we took the positions,
proper motions, radial
velocities of stars and clusters at their nominal values from the sources
described in Sec.~\ref{obs}. However,
taking into account the quality of the input data, we allowed
variations of initial conditions for O-stars in eqs. (2) and (3) 
within the given error budget.
We varied the nominal values of their proper motions and
radial velocities between -2.5~$\sigma$ and +2.5~$\sigma$ in 0.5~$\sigma$
steps,
and the distance moduli between $[(V-M_V) - 2]$ and $[(V-M_V) + 2]$ in
steps of 0.02~mag.
Initial conditions for the clusters were not varied.


The orbits were traced back in time over 11~Myr with a step
of 0.05 million years. At each time step, the relative distance
between each star and each cluster
was determined.
For the rest of the paper we adopt the following terminology.
We speak of an {\em encounter}, if, going backward in time, the distance between
star and cluster (centre) was less than 10~pc. This is called an {\em acceptable solution}.
The {\em encounter time}, $t_{enc}$, is the time before present when an 
encounter occurred, and the {\em encounter distance}, $d_{enc}$, is the distance between
star and cluster centre at $t_{enc}$. Note, however, that
the actual physical process is an {\em ejection} of a star from its cluster
counterpart. This ejection occurred at time $t_{enc}$, when the star was
at a distance $d_{enc}$ from the cluster centre.


For each acceptable solution, we computed a probability $p_{kin}$ describing
how well the proper motions and radial velocities (actually used in the solution)
suit the nominal values of the kinematic parameters of given star and cluster:
\begin{equation}
p_{kin} = \exp \left\{-\frac{1}{6}\left[\left(\frac{\Delta \mu_{\alpha,s}}{\Sigma_{\mu_{\alpha,s,c}}}\right)^2 + \left(\frac{\Delta \mu_{\delta,s}}{\Sigma_{\mu_{\delta,s,c}}}\right)^2 + \left(\frac{\Delta RV_s}{\Sigma_{RV,s,c}}\right)^2\right]\right\}.
\end{equation}
Here $\Delta \mu_{\alpha,s}$, $\Delta \mu_{\delta,s}$, and $\Delta RV_s$ are differences
between the applied and the nominal values of the corresponding velocity components of the star, and
$(\Sigma_{vc,s,c})^2 = \sigma_{vc,s}^2 + \sigma_{vc,c}^2$ where $\sigma_{vc,s}$ and
$\sigma_{vc,c}$ are the $rms$ errors of the corresponding velocity component
of the star and cluster under consideration. If $\Delta$ is smaller
than $\Sigma$ in each of three velocity components, then the corresponding
solutions has a kinematic probability $p_{kin} > 0.606$.

This procedure yielded a three-parameter set i.e., encounter time $t_{enc}$, 
present-day (or dynamic) distance $d_{dyn}$ of an O-star from the Sun, and
kinematic probability $p_{kin}$ of acceptable solutions for the encounter 
distance $d_{enc}$. In the next step, we selected  10\% of solutions with the
largest $p_{kin}$ and gave the highest priority to the solution providing the smallest
$d_{enc}$. If, for a given O-star there were solutions hitting more
than one cluster, we took into account the available information
on distance estimates for this star (e.g., spectroscopic distance,
spectro-photometric parallax, Hipparcos parallax) for making the decision. 

\begin{figure}
\resizebox{\hsize}{!}{
\includegraphics[bb=76 36 550 761,width=9cm,clip,angle=270]{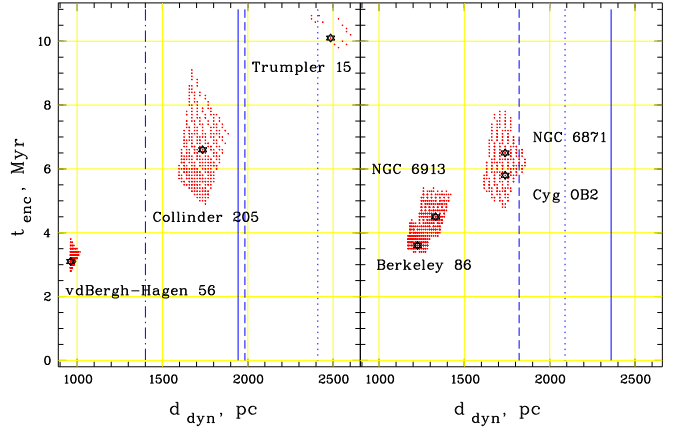}}
\caption{Dynamic distance versus back-tracing time for HD~75222 (left panel) and
HD~201345 (right panel). Dots indicate all acceptable solutions, stars show the best
solutions for an individual star-cluster combination. Vertical lines mark
different distance estimates for these stars: the solid line is for 
a spectro-photometric distance from \citet{1980A&AS...42..251N}, the dotted and
dashed lines are for spectroscopic distances based on spectral types from
\citet{2007astro.ph..3005S} and $M_V$--spectral-type calibrations from \citet{schmidtkaler}
and \citet{1992A&AS...94..211G}, respectively. The dashed-dotted line is a distance estimate
taking the spectral type from \citet{1999mctd.book.....H} and calibrations from \citet{schmidtkaler}.
}\label{fig:solu}
\end{figure}

In Fig.~\ref{fig:solu} we give two  examples of such sets of solutions. The left panel of
Fig.~\ref{fig:solu} shows the case of HD~75222, the right panel is for HD~201345. For 
HD~75222 we obtain three sets of acceptable solutions depending on the present-day distance $d_{dyn}$
of the star from the Sun. Nevertheless, the solution including vdBergh-Hagen~56 as
a cluster which the star encountered about 3~Myr ago can be rejected due to a bad
compatibility with spectroscopic and spectro-photometric distance estimates.
The solution with Trumpler~15 fits a distance estimate based on $M_V$--spectral-type 
calibrations from \citet{schmidtkaler} but needs strong variations in all three
kinematic parameters, so even for the best solution the kinematic probability is 
smaller than 0.2. On the other hand, the solution with Collinder~205 coincides better with
the other distance estimates available for HD~75222, and it is quite a stable one. We conclude: assuming a
present-day distance $d_{dyn} = 1735$~pc, HD~75222 was ejected from the
young cluster Collinder~205 at $d_{enc} \approx 1$~pc about 6.6~Myr ago. The
kinematic probability of this solution is 0.96.  

The case of HD~201345 is less clear. There are four sets of acceptable solutions.
Again, encounters with NGC~6913 and Berkeley~86 can be rejected since they
assume a present-day distance of the star of about 1.2~kpc which is too small with respect to
the spectroscopic distance estimates. The solutions with NGC~6871 and Cyg~OB2 as
counterparts assume both a present-day distance of $d_{dyn} = 1740$~pc for HD~201345
and fit much better the spectroscopic distance estimates. With a kinematic
probability $p_{kin} = 0.99$ the star was ejected from Cyg~OB2 at $d_{enc} = 1.5$~pc
about 5.8~Myr ago. For the solution with NGC~6871 we obtained $d_{enc} = 1$~pc,
$t_{enc} = 6.5$~Myr, $p_{kin} = 0.85$. Since the kinematic probability of the
solution with Cyg~OB2 is higher, we selected Cyg~OB2 as the most probable host 
of HD~201345. However, the solution with NGC~6871 cannot be rejected completely.
We discuss this case below in Sec.~\ref{indstars} which is devoted to the results on individual stars.
 
\begin{figure}
\resizebox{\hsize}{!}{
\includegraphics[bb=80 43 528 685,width=9cm,clip,angle=270]{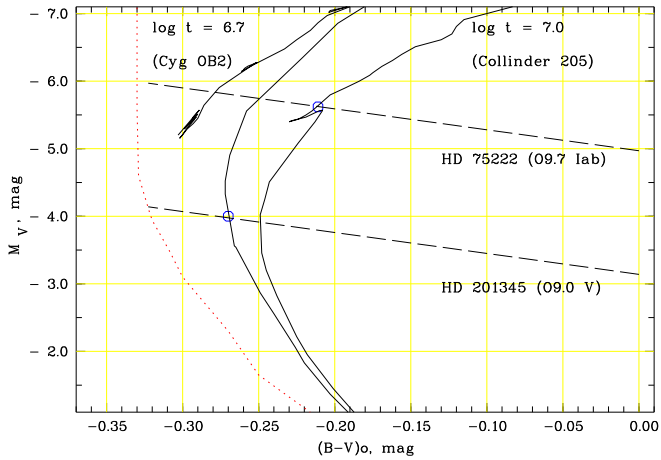}}
\caption{The colour-magnitude diagram ($(B-V)_{0}$, $M_V$) of the acceptable 
solutions for HD~75222 and
HD~201345. 
The dashed lines show $(M_V - 3.1(B-V)_{0})$ lines based on eq.~(1) and dynamical
distances $d_{dyn}$ derived from the back-tracing procedure for HD~75222 and
HD~201345. The dotted line stands for the ZAMS, whereas solid lines indicate isochrones
of Collinder~205 and Cyg~OB2 found to be the hosts of HD~75222 and
HD~201345, respectively. Small circles mark the absolute magnitude adopted
for HD~75222 ($M_V = -5.6$) and HD~201345 ($M_V = -4.0$).  
}\label{fig:mv_iso}
\end{figure}
For each star with a present-day distance $d_{dyn}$ determined, we can 
estimate its absolute magnitude $M_V$ using eq.~(1) which directly provides a relation
$(M_V - 3.1(B-V)_0)$. Assuming that the adopted extinction law is correct and
the intrinsic colour $(B-V)_0$ of an O-star is between $-0.33$~mag and $-0.13$~mag,
we obtain a maximum variation of about 0.6~mag for its absolute
magnitude $M_V$. The limit of $-0.33$~mag is defined by the location of the ZAMS, whereas
the limit $-0.13$~mag corresponds to spectral types from B7~V to B3~I \citep[see][]{schmidtkaler}
and it was chosen to be on safe ground not to exclude O-stars.  Assuming further that 
the star is ``genetically'' related to
its counterpart-cluster, we chose the crossing point between the line 
$(M_V - 3.1(B-V)_0)$ for the star and the isochrone corresponding to the
cluster age to be the absolute magnitude $M_V$ of the star. For illustration, we show
the corresponding $((B-V)_0, M_V)$ plot in Fig.~3 for HD~75222 and HD~201345.
The adopted absolute magnitudes $M_V$ are $-5.6$ and $-4.0$ for HD~75222 and HD~201345,
respectively.

\section{Results}~\label{res}

For 73 out of 93 O-stars considered, we found acceptable solutions indicating that
the present-day data are consistent with the assumption that these O-stars had encountered
(actually, are ejected from) young open clusters during the past 10 Myr.
The essential results of the back-tracing calculations are compiled in Table~\ref{tab:results}.
For each of the 73 O-stars we give: HD identification (column~1), its spectral classification
taken from the GOSV2 catalogue (2), name and age \citep{2005A&A...438.1163K,2005A&A...440..403K}
of the probable counterpart cluster (3,4),
the dynamical distance of the star $d_{dyn}$ (5). Column 6 gives the time $t_{enc}$, i.e.
the time before present when the star was ejected. Columns 7 and 8 are the distance $d_{enc}$ and the 
relative velocity $\Delta Vel_{enc}$ between the star and the cluster at $t_{enc}$. Column 9 contains
the star's absolute magnitude and its upper and lower limits estimated from $d_{dyn}$ (see 
the end of Sec.~\ref{proc}), and column 10 gives the kinematic probability $p_{kin}$ of the solution.

\begin{figure}
\resizebox{\hsize}{!}{
\includegraphics[bb=100 45 525 775,width=9cm,clip,angle=270]{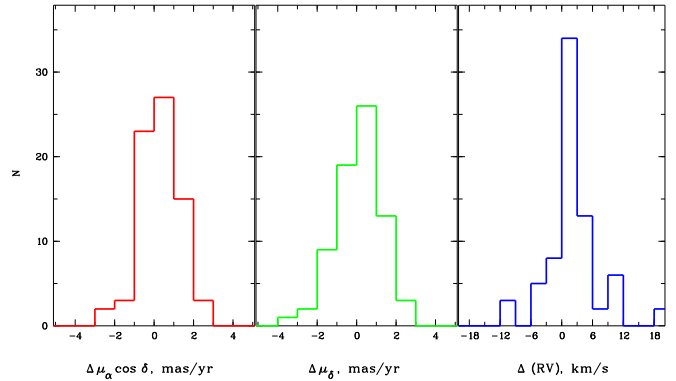}}
\caption{Distribution of off-sets introduced in proper motions $\mu_{\alpha}\cos \delta$ (left),
$\mu_{\delta}$ (middle), and in radial velocities $RV$ (right) of the O-stars
getting acceptable solutions.
}\label{fig:offsets}
\end{figure}

Except in one case, the kinematic probability $p_{kin}$ is always larger than 0.5. This
means that in units of $rms$ errors, only small variations of the kinematic parameters
were needed to get an acceptable solution for the majority of stars. The distributions of
off-sets in the sense (used parameter $-$ nominal parameter) are shown in Fig.~\ref{fig:offsets} for proper motions and 
radial velocities of the 73 stars with acceptable solutions. The distributions do not indicate
any anomalies, so we conclude that the results of the back-tracing calculations can
be used to understand where the field O-stars came from and explain their present location.

Of 161 young open clusters in our sample, only 
48 clusters were ``hosts'' of field O-stars. Only two clusters are older than 20 Myr.
Twelve clusters ejected two
O-stars each,  two clusters (Cyg OB2, and ASCC 8) had
three ejections in the past, and  three clusters (Trumpler 14, Loden~821 and ASCC~9)
even four ejections.
Three O-stars turned out to be members of the newly detected open clusters ASCC~45 and ASCC~79
\citep{2005A&A...440..403K}. Since their origin seems to be clear, they are excluded
from the statistics presented below.

In Fig.~\ref{fig:enctime} we show the ages of the clusters versus encounter time $t_{enc}$.
Close to the bisector in this figure, one expects to find O-stars which left the 
cluster birth-places just before, during or just after the time the clusters were forming.
In this sense they were members of protoclusters but did not become actual cluster members;
they  were formed in the same regions where the corresponding clusters originated.
On the other hand, there is a number of O-stars with $t_{enc}$ significantly smaller than
the ages $t_{ocl}$ of their counterparts. For these stars, we assume that they were
ejected from the already formed cluster at an early stage of the cluster's life, either due to internal 
evolutionary processes in the cluster itself, due to binary evolution or
due to external disturbing forces \citep[see e.g.][]{2002MNRAS.336.1188K,2003IAUS..212...80Z}.
We have marked these candidates by crosses in  Fig.~\ref{fig:enctime} when
$t_{enc} < 0.5 \times t_{ocl}$. This criterion is related to the accuracy of the age estimates
of open clusters in our sample which is found to be about $\sigma_{\log t} = 0.20 ... 0.25$ 
\citep{2005A&A...438.1163K}. We remark that this determination of cluster ages is based 
on the Padova isochrones \citep{2002A&A...391..195G} which have a lower limit at $\log t_{ocl} = 6.6$.
Therefore, the ages of the youngest clusters can be somewhat overestimated. Taking this
into account, the portion of ejected O-stars from already formed clusters should be about $25-35\%$.
The majority
of O-stars, however, was ejected from the star formation region during the protocluster
phase. The fact, that we do presently observe the outcome of this latter scenario as an ''isolated'' O-star and a 
{\em surviving} open cluster,
may indicate that cluster disruption by O-stars in early stages is possibly less effective than
assumed by e.g. \citet{2003ARA&A..41...57L}.
\begin{figure}
\resizebox{\hsize}{!}{
\includegraphics[bb=130 58 538 744,width=9cm,clip,angle=270]{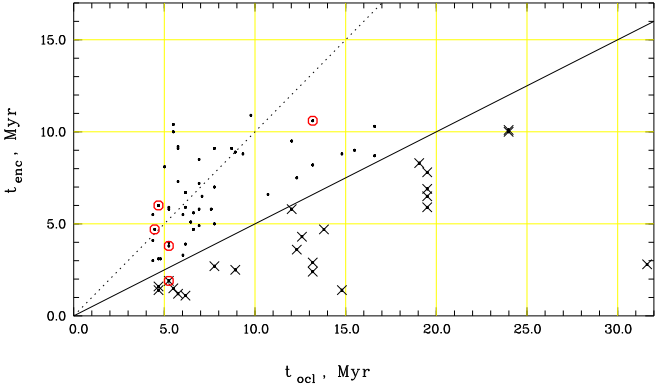}}
\caption{Cluster age $t_{ocl}$ versus encounter (ejection) time $t_{enc}$. Dots stand for the 70 O-stars with 
acceptable solutions
from the back-tracing procedure. Crosses mark 23 O-stars which were possible cluster members at
the beginning of their life but, later, were ejected from the parent clusters. The solid line
$t_{enc} = 0.5 \times t_{ocl}$ separates them from the other O-stars whereas the dotted line
is the bisector. 
The red open circles indicate O-stars with velocities larger than 80~km/s with respect to 
their cluster counterparts at the moment of the closest approach (ejection).  
}\label{fig:enctime}
\end{figure}

From the study of the initial mass function of Galactic open clusters, \citet{Pisk2008b}
found that a typical cluster
loses about $60 - 80\%$ of its initial mass during the first 260 Myrs of its evolution. 
The average mass loss rate determined by \citet{Pisk2008b} ranges from 3 to
14~$M_\odot$/Myr which includes mass loss due to stellar and dynamical evolution.
From the number of the ejected O-stars, the
number of the parent clusters, and the distribution of their encounter times $t_{enc}$,
we can roughly estimate an average mass loss of a cluster caused by the ejection of O-stars
alone. Assuming a typical mass of an O-star of 20~$M_\odot$, we obtain the average
mass loss rate of a cluster due to ejected O-stars from about 5~$M_\odot$/Myr if
$t_{enc} < 5$~Myr to 3~$M_\odot$/Myr if 5~Myr$ < t_{enc} < 11$~Myr. Though the statistics is rather poor,
the result agrees well with the estimation by \citet{Pisk2008b} and underlines the
importance of the contribution of ejected O-stars to the general mass loss of
open clusters in the first 10~Myr of their life.

\begin{figure}
\resizebox{\hsize}{!}{
\includegraphics[bb=99 58 542 680,width=9cm,clip,angle=270]{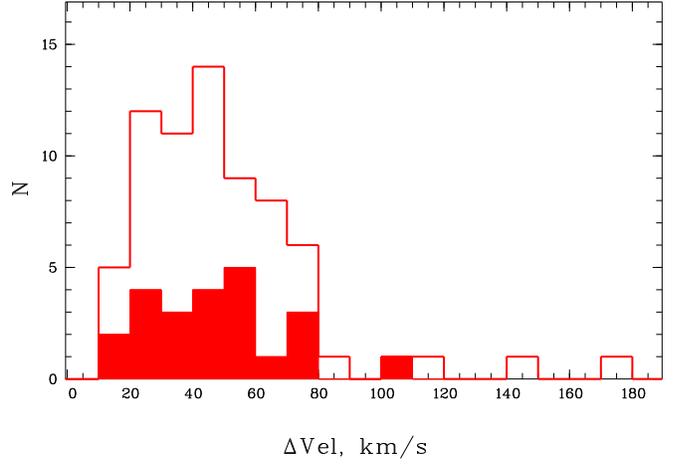}}
\caption{Distribution of relative velocities of O-stars with respect to their cluster
counterparts at the moment of the closest approach (ejection). The filled histogram shows
former cluster members ejected from their hosts.}\label{fig:relvel}
\end{figure}

At the moment of ejection, the relative velocities of former 
cluster members with respect to their parent
clusters are rather moderate, with median  at $\Delta Vel \approx 45$~km/s. This
does not change significantly if we consider all 70 O-stars. In Fig~\ref{fig:relvel} we show the histogram of
relative velocities $\Delta Vel$ for all O-stars with acceptable solutions and for the 
23 stars  probably having been cluster members in the past. Of five stars with $\Delta Vel$
larger than 80~km/s, there is only one which was a former cluster member.  The other four
belonged probably to protoclusters and  were ejected from the regions during cluster
formation.

\begin{figure}
\resizebox{\hsize}{!}{
\includegraphics[bb=133 44 502 762,width=9cm,clip,angle=270]{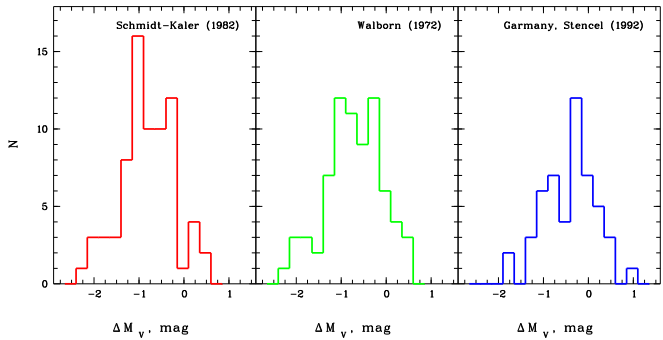}}
\caption{Distribution of differences in the absolute magnitudes of field O-stars
derived from our dynamical $d_{dyn}$ distances versus the
absolute magnitudes according to the calibrations by \citet{schmidtkaler},
\citet{1972AJ.....77..312W}, and \citet{1992A&AS...94..211G}, respectively.
On average, our $d_{dyn}$ lead to systematically lower
$M_V$ than the previous  calibrations.}\label{fig:diff_abs_mag}
\end{figure}

As described in Sec.~4, our back-tracing procedure allows to determine the present-day,
or dynamical distances $d_{dyn}$ of O-stars. This gives us estimates of their absolute magnitudes $M_{V,dyn}$.
In Fig.~\ref{fig:diff_abs_mag} we show the distribution of differences $\Delta M_V = M_{V,st} - M_{V,dyn}$ to
the standard calibrations from \citet{schmidtkaler},
\citet{1972AJ.....77..312W}, and \citet{1992A&AS...94..211G}.
At first glance this picture is intriguing, but remember that the calibrations of 
absolute magnitudes of O-stars date back to the pre-Hipparcos era. 
After Hipparcos it has become possible to start a re-calibration, admittedly more for early B-type stars
than for the O-stars themselves. Utilizing the measurements from Hipparcos,
\citet{1999ASPC..167..263K} found 
$M_{V,st} - M_{V,Hipp} = -0.85 \pm 0.12 $ for 44 B0-B3 III, IV  stars. This number coincides well with our findings 
for the median of $\Delta M_V = -0.84, -0.72$ or $-0.36$ depending on the different calibrations (see Fig.~\ref{fig:diff_abs_mag}).
As we will show below, $\zeta$ Pup, the only O-star with a highly significant 
parallax in \citet{2007hnrr.book.....V}, has 
an $M_{V,st} - M_{V,Hipp} = -1.34$. 

\section{Individual stars}~\label{indstars}

In the following we discuss the results for a few selected stars in
more detail. These include the case of $\zeta$ Pup, the closest O-star to
the Sun, a case of a common origin of 4 O-stars, and finally a case of O-stars
presently at large distance from the Galactic plane.

\subsection{$\zeta$ Pup (HD~66811)}

$\zeta$ Pup is  an important benchmark for the astrophysical characteristics 
of massive stars. Spectroscopically it is classified as
O4\,I according to \citet{2007astro.ph..3005S}, and its trigonometric
parallax from the original Hipparcos catalogue \citep{1997yCat.1239....0E} is  
2.33$\pm$0.51 mas.  From a back-tracing of $\zeta$ Pup, \citet{2001A&A...365...49H}
found that this star had a possible encounter with the cluster Trumpler~10
some 2~Myr ago provided that its dynamical distance was $d_{dyn} = 250$ ... 300~pc.
This was inconsistent with the Hipparcos distance of 430~pc, and also 
its absolute magnitude did not agree with the cluster isochrone.
Our back-tracing confirms the results from \citet{2001A&A...365...49H},
giving Trumpler~10 as the host and $d_{dyn} = 300$~pc,
$t_{enc} = 2.5$~Myr, $d_{enc} = 0.9$~pc, and $p_{kin} = 0.94$. This result 
is consistent with the new Hipparcos parallax (3.00$\pm$0.1 mas) from the 
re-reduction of Hipparcos data by \citet{2007hnrr.book.....V}.
If we literally adopt the new Hipparcos distance of 333~pc, we get a solution with
$t_{enc} = 1.8$~Myr and $d_{enc} = 7.1$~pc, which has only a slightly smaller
probability ($p_{kin} = 0.91$).

In Table \ref{tabzetpup}  we summarise the 
distances of $\zeta$ Pup from the various sources. Compared to the
new Hipparcos benchmark, the old spectroscopic distance renders
$\zeta$ Pup 1.34 mag too bright in absolute magnitude. According to our 
$d_{dyn}$ it would be 0.24 mag fainter than based on the revised Hipparcos distance.
From a  non-LTE analysis of the spectrum, \citet{1983A&A...118..245K}
found that the effective temperature of $\zeta$ Pup is T$_{eff}$ = 42000~K
instead of 50000~K according to spectral type O4.
This, together with the low $\log g = 3.5$, means that $\zeta$ Pup is
already away from the ZAMS.
\begin{table}
\caption{$\zeta$ Puppis: distances, distance moduli $V-M_V$, and
derived differences in absolute magnitude M$_V$ with respect to
the new Hipparcos parallax \citep{2007hnrr.book.....V}.
}             %
\label{tabzetpup}      
\centering                          
\begin{tabular}{c c r c}        
\hline\hline                 
distance, pc & $V-M_V$ & $\Delta M_V$ & sources for distance\\    
\hline                        
   333 & 7.60  & 0.00 & \citet{2007hnrr.book.....V} \\     
   429 & 8.16  & $-0.56$ & \citet{1997yCat.1239....0E} \\
   615 & 8.94  & $-1.34$ & Spectral type O4 I \\
   300 & 7.36  & $+0.24$ & this paper \\
\hline                                   
\end{tabular}
\end{table}

\citet{1983A&A...118..245K} also determined the angular diameter of $\zeta$ Pup 
to be $\alpha = 4.0\times10^{-4}$ arcsec. Together with an assumed distance of
(450$\pm$ 200)~pc, this yielded a radius of $(19 \pm 8) R_\odot$ and a mass of 
40~$M_\odot$.  Using the new Hipparcos parallax, we
find a radius of $(14\pm 0.4)R_\odot$ and a mass of (22.5$\pm$1.3)~$M_\odot$.
For its luminosity we find log$L/L_\odot = 5.74\pm0.02$. For the error
calculation, only the mean error of the Hipparcos parallax is considered.
The new Hipparcos parallax rules out the scenarios by
\citet{1996A&A...305..825V}. They discussed an origin of $\zeta$ Pup in 
Vela R2, which would imply a present-day distance of the star of 700 to 800 pc.
The alternative scenario that $\zeta$ Pup originated as a field star and
its runaway nature is due to a binary history is now also ruled out, because 
this would lead to a present-day distance between 400 and 800 pc \citep{1996A&A...305..825V}.

To summarise, $\zeta$ Pup is closer, less massive, and less luminous than
previously thought.

\subsection{HD~188209, HD~189957, HD~198846, HD~201345}

\begin{figure}
\resizebox{\hsize}{!}{
\includegraphics[bb=40 64 558 701,width=9cm,clip,angle=270]{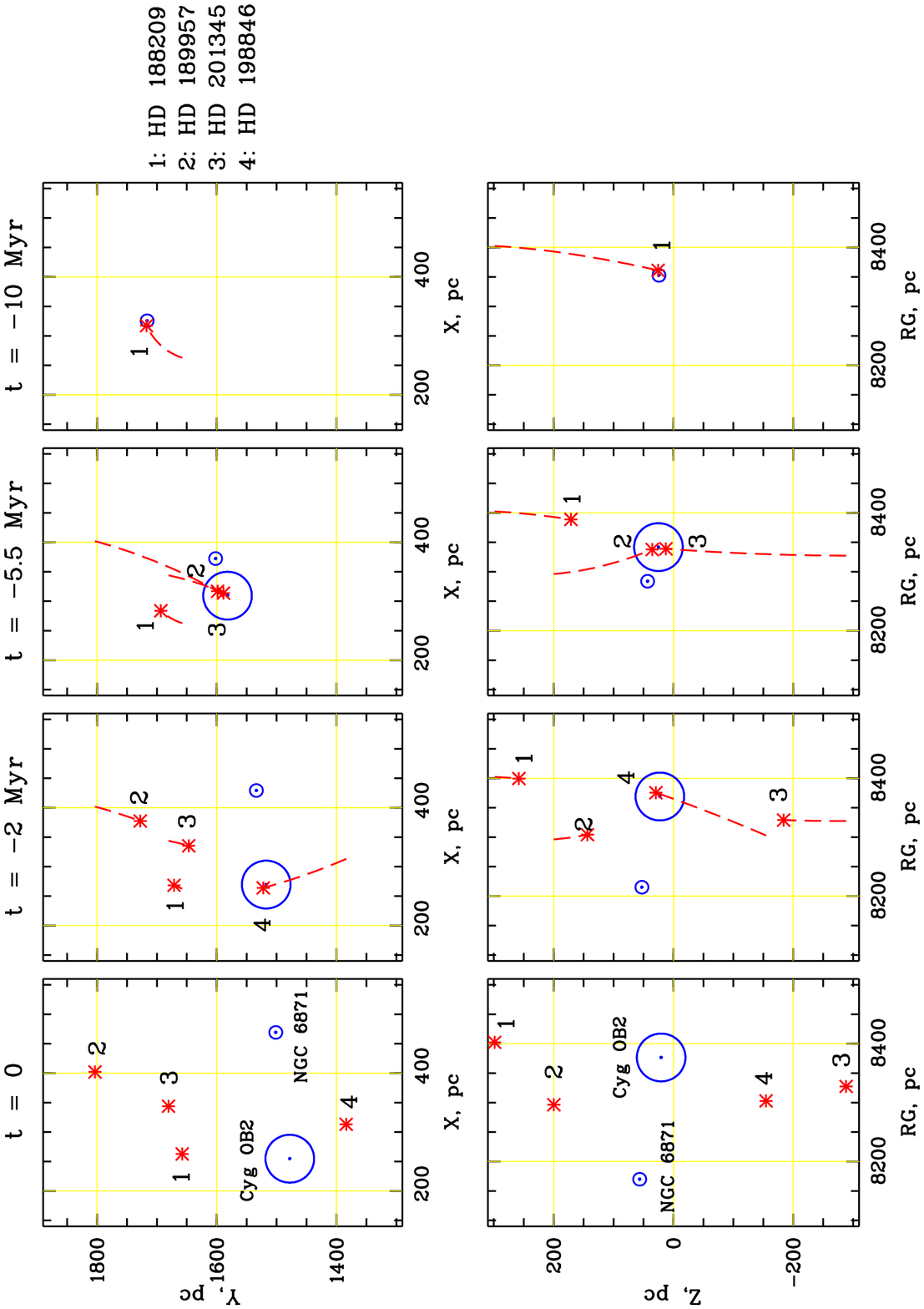}}
\caption{Spatial distribution of the O-stars (1: HD~188209, 2:  HD~189957, 3: HD~201345,
4: HD~198846), of the open cluster NGC~6871, and of the association Cyg~OB~2. The upper
panels are for the $(X,Y)$-plane, the lower panels show the distribution in the $(RG,Z)$-plane,
where RG is the Galactocentric radius. Asterisks mark the location of stars, the circles represent the
clusters; their radii are defined as the {\it present-day} tidal radii determined 
in \citet{2008A&A...477..165P}. The left columns show the location at present;
the second, third, and forth columns show the locations 2~Myr, 5.5~Myr, and 10~Myr ago,
respectively. The dashed curves in these panels delineate the stellar orbits computed with 
the back-tracing procedure from present to the corresponding time $t$ indicated on the top.
Note that Cyg~OB~2 is younger than 10~Myr, and therefore does not appear in the 
right-most column. 
}\label{fig:example2}
\end{figure}

Although, at present, these stars are separated by hundreds of parsecs, they turn out to have
common ``relatives''. We show their spatial distribution in Fig.~\ref{fig:example2} where $X$, $Y$, $Z$ are
the Cartesian Galactic coordinates and $RG$(pc) is the Galactocentric radius of a star or a cluster.
The present-day location is shown in
the left column ($t = 0$), whereas the right column ($t = -10$~Myr) displays the same region
10~Myr ago. 
At that time the open cluster NGC~6871 had formed \citep[its age is dated  $\log t = 6.99$
in][]{2005A&A...438.1163K}. Just before, at $t_{enc} = 10.9$~Myr,
the O-star HD~188209 escaped from this region with a relative velocity of about 35~km/s,
away from the Galactic plane, and towards the North pole. A few million years later
($t = - 7.7$~Myr, not shown in Fig.~\ref{fig:example2}), NGC~6871 passed at about 35~pc the region where
the association Cyg~OB~2 came into being at $t \approx -5.5$~Myr. \citet{2005A&A...438.1163K} give $\log t = 6.72$ for Cyg~OB~2.
Immediately before, two O-stars were ejected from this region,
HD~189957 ($t_{enc} = 5.9$~Myr) and HD~201345 ($t_{enc} = 5.8$~Myr). HD~189957 started
in the direction to the North pole with a relative velocity of about 70~km/s whereas
HD~201345 moved to the South pole with about the same relative velocity. About 3.5~Myr
later ($t = - 2$~Myr) the O-star HD~198846 was ejected from Cyg~OB~2 with a relative velocity
higher than 100~km/s in the direction to the South pole, too.

As we note in Sec.~3 and show in Fig.~\ref{fig:solu}, there is, although with a slightly smaller kinematic
probability, another acceptable solution for
HD~201345 suggesting an ejection from NGC~6871 about 6.5~Myr ago. Also for HD~189957,
a second acceptable solution turned out to be possible (an ejection from NGC~6871 with the parameters
$d_{dyn} = 1964$~pc, $t_{enc} = 7.1$~Myr,
$d_{enc} = 3.5$~pc, $p_{kin} = 0.84$). Since the time of their closest approach with
NGC~6871 is comparable with the time when NGC~6871 passed the birth place of Cyg~OB~2,
one may suspect that this event was the reason for the ejection of HD~201345 and HD~189957
from a region in between NGC~6871 and Cyg~OB~2.

\subsection{O-stars at large distances from the Galactic plane}

The majority of O-stars in our sample is located within 200~pc from
the Galactic plane. However, there are five stars with a present-day location at
$|Z| >$ 400~pc. These stars either had relative velocities at the moment of encounter (ejection) larger than
100~km/s (HD~116852, HD~157857) or they left their cluster counterparts more
than 9~Myr ago (HD~14633, HD~105056, HD~175754). Below, we consider two
examples, HD~116852 and HD~14633, which have the largest distances from the
Galactic plane. These cases are also interesting since their counterparts were
clusters hosting other field O-stars, too.

\subsubsection{HD~116852 and Trumpler~14}
Judging from the spectroscopic distance and high galactic latitude of HD~116852,
one can expect that this star is located at a relatively large distance from the Galactic plane.
Indeed, from back-tracing we obtain  $Z \approx -690$~pc for the present-day location of HD~116852.
This star left its birth-place in a protocluster of Trumpler~14 with a relative velocity of $\approx 180$~km/s
about 6~Myr ago, i.e. just before
the cluster was formed. \citet{2005A&A...438.1163K} list $\log t = 6.67$ for Trumpler~14.
The location of  the star and of the protocluster at that time was at $Z \approx -50$~pc.
During the next 6~Myr, the cluster moved by only 60~pc, whereas HD~116852 did more than 1~kpc.
Merely 1.5~Myr after its birth, Trumpler~14 lost another O-star, HD~93652, and
about 1.5~Myr later, HD~91651 and HD~305539 left the cluster, too. Presently, HD~93652,
HD~91651, and HD~305539 are located about 40~pc, 75~pc, and 30~pc away from Trumpler~14,
respectively. According to \citet{1995AstL...21...10M}, there is an OB-association Car~1~F at about
65~pc from Trumpler~14. Probably, this neighbourhood has had impact on the fate of
Trumpler~14.

\subsubsection{HD~14633 and ASCC~8}
Though  presently HD~14633 is at $Z \approx -670$~pc, about 9~Myr ago this star was located  
near the Galactic plane between the clusters
NGC~869 (h Per) and NGC~884 ($\chi$ Per), i.e. within a region wellknown as the then
very active star formation region Per~OB~1. About 3.5~Myr later this place was also 
the birth-place of an open cluster, ASCC~8 ($\log t = 6.76$). 
In Fig.~9 we show the location  of these three clusters (NGC~869, NGC~884, ASCC~8) as it is now and as
it was in the past. 
\begin{figure}
\resizebox{\hsize}{!}{
\includegraphics[bb=142 34 498 786,width=9cm,clip,angle=270]{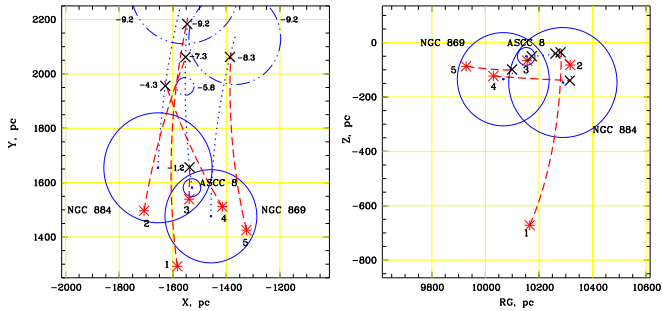}}
\caption{Spatial distribution of the field O-stars (1: HD~14633,  2:  HD~17603,  3: HD~14947, 
4: HD~12993, 5: HD~13022), and of the open clusters  NGC~869, NGC~884, ASCC~8.
The left panel is for the $(X,Y)$-plane, whereas right panel shows the distribution in the $(RG,Z)$-plane,
RG is the Galactocentric radius. Asterisks mark the present-day location of the stars, the circles are 
for the clusters: solid curves for the present-day location and  dash-dot-dot-curves for
the past. Their radii are defined by the {\it present-day} tidal radii determined
in \citet{2008A&A...477..165P}. The dashed and dotted curves show the
orbits of the stars and clusters, respectively, computed with the back-tracing procedure.
The crosses mark the places where the stars encountered their hosts. Negative
numbers in the left panel express the time in Myr when the corresponding events happened
(in order not to overload the figure, the numbers are omitted in the right panel). 
}\label{fig:example4}
\end{figure}

According to the tidal radii determined in \citet{2008A&A...477..165P}, NGC~869 and NGC~884 are
the largest clusters in the whole COCD cluster sample. Moreover, their tidal spheres are overlapping, 
and some concentration of stars above the background is observed in this overlapping zone. 
Based on common proper motions, an open cluster (ASCC~8) was identified here \citep{2005A&A...440..403K},
with an age of about 5.8~Myr. Obviously, such a location of a cluster 
does not present a good chance for a long life. From the back-tracing computations, three
O-stars marked as field in the GOSV2 came from the region occupied now by ASCC~8:
HD~14947 - 1.2~Myr ago, HD~17603 - 7.3~Myr ago, and HD~14633 - 9.2~Myr ago. Two other
field stars, HD~12993 and HD~13022, had their origins in NGC~884 and  NGC~869, respectively.
Note that the final solution for a given star and its host is selected from all acceptable
solutions ($d_{enc} < 10$~pc) to have the smallest encounter distance $d_{enc}$ at the 
highest kinematic probability $p_{kin}$. This approach is justified if possible hosts 
(i.e. open clusters) are separated by distances which are considerably larger than their sizes.
This is the usual case. However, the example considered presents an exception from
the rule. Here the distances between cluster centres are comparable to their tidal radii.
Therefore, solutions with $d_{enc} > 10$~pc can be considered, too. For example,
HD~12993 could have been  ejected from NGC~869 about 0.8~Myr ago when the star was $\approx 50$~pc
away from the cluster centre. In this special case, we prefer a more general statement:
the observational data for these five field O-stars are consistent with the claim that
they originated in Per~OB~1.

\section{Stars without solutions}~\label{nosolu}

For 73 out of 105 O-stars with assignments as $field$ stars in the GOSV2
we could trace back their origins and found the corresponding host clusters or protoclusters. 
Twelve stars could not be treated by our method, because no radial velocities were available for them.
For another twenty, we could not find a solution, i.e. they could not be associated
with any of the clusters of our sample in the past 10 million years. This negative finding,
however, should not be interpreted as a proof that these O-stars have formed in isolation.

We could not find a parameter or parameters which distinguish these stars from the other field O-stars  
which had an acceptable solution. Both groups cover a similar range of apparent magnitudes, and they
are comparable in the distribution of the mean errors of their kinematic components.
Of course, one cannot exclude that a few of them have a true distance modulus differing 
from the spectroscopic
estimates by more than two magnitudes, and/or their true velocity components are 
outside the intervals checked with the backward procedure. Nevertheless, a more 
important aspect seems to be that our sample of potential 
host clusters is not complete for associations and very young (embedded) clusters, as we stressed
in Sec.~2.1. Provided that this assumption is correct, one would expect about 20\% of field O-stars
having their origins in  these kinds of objects. To answer the question with certainty, however, more
accurate data on the distances and kinematics of associations and embedded clusters are required.

\section{Summary}~\label{conclus}

In this paper we have followed the dynamical history of O-stars that left the
groups where they originated in. We leave open the physical mechanisms which are behind
these events.

For 73 out of 93 O-stars considered, we found acceptable solutions indicating that
the present-day data are consistent with the assumption that these O-stars were ejected 
from young open clusters or protoclusters during the past 10 Myr.
The GOSV2 catalogue 
counts 370 O-stars, for 265 of which the origin is given in that catalogue. We were able 
to add 73 more cases to the list. For 32 stars (or 9$\%$)
we could not prove the origin in groups. 

In this paper we have dealt with aspects of the
early phases in the life      of open clusters.                      
O-stars are best suited as tracers of this early-phase evolution because of their
short life-time. In our sample of 161 young  open clusters (age $<$ 30 Myr) from
the COCD there are 55 (or $\approx$ 35\%) that have O-stars as members \citep{2005A&A...438.1163K,2005A&A...440..403K}, 
23 of these have already lost one or more O-stars in their history.
Another 24 (or 15\%) of the COCD clusters had relations to O-stars in the past 10 Myrs,
but do not contain O-stars at present.
For 82 (51\%) young clusters we cannot prove a relation to
presently living
O-stars. Either their most massive member is a main sequence star of spectral type
later than O, or it is a former O-star which has already evolved.
Of   the 47 clusters that have lost at least one O-star, we find 14 that are so young
that O-star and cluster should already have separated in the protocluster phase.

Summing up the statistics above, the following picture emerges. Fifty percent of the clusters being
able to survive the infant-mortality phase are so massive that they contain or contained O-stars.
These O-stars have not been able to destroy the cluster. This, in parts, answers
the question asked by \citet{2003ARA&A..41...57L}:
Do the progenitors of bound open clusters ever contain O-stars? Yes, they did.

The fact, that we could not trace back 9\% of all O-stars from the GOSV2, 
does not 
necessarily mean that ``isolated O-star formation'' is possible. Their known astrophysical
data (distance, velocities) may be incorrect or our list of possible host candidates may be incomplete.
On the other hand, we can interpret our result - no solution for 20 out of 93 stars - as follows:
there is an upper bound of slightly more than 20\% of O-stars which could have destroyed
their family of brothers and sisters with which they may have formed together.

It has been shown by \citet{Pisk2008b} that classical (gravitationally bound) open clusters in the 
Milky Way evolve due to stellar and dynamical evolution as well as due to 
external perturbations. They
are losing stellar mass during their live-time at an average rate of 3 to
14~$M_\odot$/Myr. In this paper
we determined the mass loss rate of young open clusters due to
O-stars alone to be  3 to 5~$M_\odot$/Myr in the first few million years of their existence.

As a by-product, we find new distances and absolute magnitudes for 73 O-stars. These indicate
that the calibration of absolute magnitudes of O-stars
should be revised. Their absolute magnitudes are systematically fainter by about
0.3 to 0.8 mag compared to the calibrations by  \citet{1992A&AS...94..211G}, \citet{1972AJ.....77..312W},
or \citet{schmidtkaler}. This would be consistent with the re-calibration of the absolute magnitudes
of early B-type stars by \citet{1999ASPC..167..263K} using Hipparcos trigonometric parallaxes.

We have also shown that $\zeta$ Pup, the closest O-star from the Sun,
left the young open cluster Trumpler 10 some  2.5 Myrs ago. Its present-day
distance from the Sun of 300 pc is compatible with the new Hipparcos distance
from \citet{2007hnrr.book.....V}. This implies a radius of 14 $R_\odot$, a mass of 22.5 $M_\odot$,
and a luminosity of log$L/L_\odot$ of 5.74 for $\zeta$ Pup, i.e. the values are considerably smaller than
assumed before.

\begin{acknowledgements}
We are grateful to Nina Kharchenko, Anatoly Piskunov and Hans Zinnecker for fruitful discussions
on massive star formation and young open clusters.

\end{acknowledgements}

\bibliographystyle{aa}
\bibliography{OSTARS}

\begin{longtable}{rllcrrrrccl}
\caption{\label{tab:results} Results of the back-tracing calculations}\\
\hline\hline
HD &Sp. type&Cluster name&$\log t_{ocl}$& $d_{dyn}$&$t_{enc}$&$d_{enc}$&$\Delta Vel_{enc}$ &$M_{V}$ [min, max]&$p_{kin}$&comments\\
   &        &      &      &[pc]     &[Myr]   & [pc]   &[km/s]& [mag]           &         &        \\
\hline
\endfirsthead
\caption{continued.}\\
\hline\hline
\hline
\endhead
       1337&O9.0 III   &NGC 957          & 6.84 &1911& 8.5& 3.9& 75.7&-6.0 [-6.4, -5.8]&0.54&                   \\
      10125&O9.7 II    &IC 1590          & 6.84 &2300& 4.9& 6.1& 78.5&-5.3 [-5.6, -5.0]&0.81&                   \\
      12323&O9.0 V     &ASCC  9          & 6.79 &2890& 6.7& 4.6& 46.8&-4.0 [-4.1, -3.5]&0.97&                   \\
      12993&O6.5 V     &NGC 884          & 7.10 &2074& 4.3& 0.3& 61.2&-4.0 [-4.2, -3.6]&0.71&                   \\
      13022&O9.5 II-III&NGC 869          & 7.28 &1948& 8.3& 2.6& 17.2&-4.3 [-4.7, -4.1]&0.88&                   \\
      13745&O9.7 II    &NGC 663          & 7.14 &2035& 4.7& 2.5& 50.5&-4.9 [-5.2, -4.6]&0.80&                   \\
      14434&O5.5 V     &ASCC  9          & 6.79 &3062& 5.9& 5.5& 34.0&-5.2 [-5.5, -4.9]&0.81&                   \\
      14442&O5.0 V     &ASCC  9          & 6.79 &2912& 3.9& 3.4& 40.9&-5.2 [-5.4, -4.8]&0.94&                   \\
      14633&O8.0 V     &ASCC  8          & 6.76 &2151& 9.2& 3.8& 76.5&-4.5 [-4.6, -4.0]&0.89&  (1)              \\
      14947&O5.0 I     &ASCC  8          & 6.76 &2177& 1.2& 8.8& 32.1&-5.8 [-6.1, -5.5]&0.77&                   \\
      15137&O9.5 II-III&NGC 957          & 6.84 &1784& 5.8& 2.4& 27.9&-4.3 [-4.5, -3.9]&0.92&                   \\
      16691&O4.0 I     &ASCC  9          & 6.79 &2901& 1.1& 6.8& 43.3&-5.8 [-6.1, -5.5]&0.55&                   \\
      17603&O7.5 Ib    &ASCC  8          & 6.76 &2272& 7.3& 1.2& 26.5&-6.1 [-6.3, -5.7]&0.97&                   \\
      39680&O6.0 V     &NGC 2169         & 6.89 &1354& 9.1& 5.5& 35.2&-3.6 [-3.8, -3.2]&0.84&                   \\
      41997&O7.5 V     &Collinder 89     & 7.50 & 723& 2.8& 2.4& 44.4&-2.7 [-3.0, -2.5]&0.63&                   \\
      44811&O7.0 V     &NGC 2129         & 7.08 &1582& 9.5& 8.9& 18.3&-3.7 [-4.0, -3.4]&0.51&                   \\
      52266&O9.0 IV    &Collinder 106    & 6.74 &1388&10.4& 1.9& 34.3&-4.3 [-4.4, -3.8]&0.55&                   \\
      52533&O9.5 V     &Collinder 106    & 6.74 &1831&10.0& 3.8& 34.9&-4.2 [-4.4, -3.7]&0.86&                   \\
      57236&O8.0 V     &NGC 2414         & 6.94 &2432& 9.1& 5.4& 62.3&-4.6 [-4.8, -4.2]&0.84&                   \\
      66811&O4.0 I     &Trumpler 10      & 6.95 & 297& 2.5& 1.4& 51.0&-5.2 [-5.3, -4.7]&0.94&  (1)              \\
      69464&O6.5 Ib    &ASCC 45          & 7.12 &3001& 0.1& 0.8& 11.4&-5.2 [-5.6, -4.9]&0.78&  (2)              \\
      75222&O9.7 Iab   &Collinder 205    & 7.03 &1735& 6.6& 1.1& 62.8&-5.6 [-6.0, -5.4]&0.97&  (1)              \\
      76968&O9.7 Ib    &ASCC 45          & 7.12 &2682&10.6& 4.1& 81.2&-5.9 [-6.5, -5.8]&0.86&                   \\
      89137&O9.5 III   &Loden 306        & 6.76 &2026& 9.1& 3.1& 45.6&-4.3 [-4.4, -3.8]&0.91&                   \\
      91651&O9.0 V     &Trumpler 14      & 6.67 &2720& 1.6& 2.3& 46.3&-4.2 [-4.3, -3.7]&0.91&                   \\
      93632&O5.0 III   &Trumpler 14      & 6.67 &2731& 3.1& 6.0& 12.3&-5.6 [-5.8, -5.1]&0.74&                   \\
      94963&O6.5 III   &IC 2581          & 7.22 &2504& 8.7& 6.5& 27.8&-5.1 [-5.6, -5.0]&0.81&                   \\
      96917&O8.5 Ib    &Collinder 228    & 6.68 &1991& 3.1& 1.9& 55.5&-5.5 [-5.7, -5.1]&0.84&                   \\
      96946&O6.0 V     &vdBergh-Hagen 121& 6.64 &2738& 4.1& 7.8& 49.2&-5.3 [-5.4, -4.8]&0.70&                   \\
      97848&O8.0 V     &NGC 3324         & 6.72 &2329& 4.0& 8.8& 48.0&-4.0 [-4.1, -3.5]&0.90&                   \\
     104565&O9.7 Ia    &ASCC 75          & 6.65 &2839& 4.7& 3.2&148.0&-5.1 [-5.2, -4.5]&0.72&                   \\
     104649&O9.5 V     &NGC 3572         & 6.88 &1948& 5.8& 2.2& 40.2&-4.4 [-4.6, -4.0]&0.93&                   \\
     105056&O9.7 Ia    &Ruprecht 94      & 7.19 &3304& 9.0& 3.8& 47.4&-5.8 [-6.4, -5.7]&0.64&  (1)              \\
     105627&O9.0 II-III&Loden 821        & 7.29 &2885& 6.5& 7.5& 70.8&-4.8 [-5.3, -4.7]&0.68&                   \\
     112244&O8.5 Iab   &Feinstein 1      & 6.97 &1391& 8.8& 3.2& 45.3&-6.0 [-6.4, -5.8]&0.83&                   \\
     116852&O9.0 III   &Trumpler 14      & 6.67 &2475& 6.0& 2.7&179.6&-4.1 [-4.2, -3.6]&0.67&  (1)              \\
     117856&O9.5 III   &Loden 694        & 7.38 &1750&10.0& 5.2& 17.4&-5.0 [-5.5, -4.9]&0.51&                   \\
     120521&O8.0 Ib    &Loden 821        & 7.29 &2890& 7.8& 1.8& 26.5&-5.0 [-5.5, -4.8]&0.80&                   \\
     120678&O8.0 III   &Loden 821        & 7.29 &2864& 5.9& 1.5& 39.8&-5.2 [-5.8, -5.1]&0.84&                   \\
     123008&O9.7 Ib    &Loden 821        & 7.29 &2875& 6.9& 1.4& 51.0&-5.1 [-5.6, -5.0]&0.66&                   \\
     123056&O9.5 V     &Loden 694        & 7.38 &1577&10.1& 3.1& 28.5&-3.9 [-4.3, -3.7]&0.83&                   \\
     125206&O9.5 IV    &NGC 5606         & 6.84 &2054& 7.2& 3.6& 41.1&-5.2 [-5.5, -4.9]&0.68&                   \\
     135240&O7.5 III   &ASCC 79          & 6.86 & 809& 2.2& 0.6&  5.2&-5.0 [-5.3, -4.7]&0.73&  (3)              \\
     135591&O7.5 III   &ASCC 79          & 6.86 & 796& 0.3& 1.4& 17.1&-4.6 [-4.8, -4.2]&0.74&  (3)              \\
     148546&O9.0 Ia    &ASCC 88          & 7.17 &1453& 8.8& 5.7& 55.4&-4.7 [-5.0, -4.4]&0.73&  (1)              \\
     153426&O9.0 II-III&Hogg 22          & 6.70 &1694& 8.1& 9.9& 55.4&-4.9 [-5.1, -4.5]&0.91&                   \\
     153919&O6.5 Ia    &NGC 6231         & 6.81 &1034& 5.1& 1.1& 48.8&-5.1 [-5.4, -4.7]&0.82&                   \\
     154368&O9.5 Iab   &Sco OB4          & 6.82 &1082& 4.7& 5.3& 11.2&-6.3 [-6.6, -6.0]&0.65&                   \\
     154643&O9.5 III   &ASCC 88          & 7.17 &1908& 1.4& 2.3& 20.3&-5.6 [-6.1, -5.5]&0.85&                   \\
     154811&O9.7 Iab   &vdBergh-Hagen 205& 7.12 &2019& 8.2& 2.7& 35.1&-6.2 [-6.8, -6.2]&0.96&                   \\
     156212&O9.7 Iab   &Trumpler 28      & 6.89 & 992& 7.0& 4.0& 53.5&-4.5 [-4.7, -4.1]&0.93&  (1)              \\
     157857&O6.5 III   &NGC 6611         & 6.72 &1854& 3.8& 2.6&113.5&-4.9 [-5.1, -4.5]&0.68&  (1)              \\
     158186&O9.5 V     &Sco OB4          & 6.82 &1254& 5.6& 1.2& 30.3&-4.4 [-4.6, -4.0]&0.82&                   \\
     161853&O8.0 V     &Trumpler 28      & 6.89 &1241& 5.0& 2.6& 25.4&-4.1 [-4.3, -3.7]&0.68&                   \\
     166734&O7.0 Ib    &NGC 6604         & 6.64 &1661& 3.0& 2.5& 22.4&-6.7 [-7.0, -6.4]&0.65&                   \\
     169515&O9.7 Ib    &Markarian 38     & 6.95 &1794& 8.9& 4.3& 37.7&-6.1 [-6.5, -5.9]&0.54&                   \\
     169582&O6.0 I     &NGC 6604         & 6.64 &1565& 5.5& 3.2& 28.5&-4.9 [-5.0, -4.4]&0.57&                   \\
     171589&O7.0 II    &NGC 6618         & 6.78 &1934& 3.3& 1.5& 63.2&-4.9 [-5.1, -4.5]&0.96&                   \\
     175754&O8.0 II    &ASCC 93          & 7.22 &2586&10.3& 4.9& 64.8&-5.3 [-5.8, -5.2]&0.47&                   \\
     175876&O6.5 III   &NGC 6618         & 6.78 &1922& 5.5& 4.8& 61.7&-5.0 [-5.2, -4.5]&0.88&                   \\
     188001&O7.5 Ia    &Roslund 2        & 6.89 &2022& 2.7& 2.5& 70.3&-6.0 [-6.4, -5.7]&0.79&  (1)              \\
     188209&O9.5 Iab   &NGC 6871         & 6.99 &1705&10.9& 7.3& 33.7&-5.9 [-6.4, -5.7]&0.57&                   \\
     189957&O9.5 III   &Cyg OB2          & 6.72 &1858& 5.9& 3.1& 67.7&-4.4 [-4.6, -4.0]&0.87&  (1)              \\
     192281&O5.0 V     &NGC 6913         & 7.12 &1081& 2.4& 1.2& 35.5&-4.5 [-4.8, -4.2]&0.80&                   \\
     195592&O9.7 Ia    &NGC 6913         & 7.12 &1132& 2.9& 6.3& 40.3&-6.3 [-6.9, -6.3]&0.79&                   \\
     198846&O9.0 V     &Cyg OB2          & 6.72 &1427& 1.9& 3.7&106.6&-4.1 [-4.3, -3.7]&0.81&  (1)              \\
     201345&O9.0 V     &Cyg OB2          & 6.72 &1739& 5.8& 1.5& 65.6&-4.0 [-4.1, -3.5]&0.99&                   \\
     203064&O7.5 III   &Collinder 419    & 6.85 & 631& 6.5& 1.3& 26.9&-4.8 [-5.0, -4.4]&0.87&  (1)              \\
     218915&O9.5 Iab   &ASCC 120         & 7.08 &2369& 5.8& 4.3& 54.7&-5.4 [-5.8, -5.1]&0.72&  (1)              \\
     303492&O9.0 Ia    &Loden 153        & 6.74 &2714& 1.5& 2.6& 79.3&-5.8 [-6.1, -5.5]&0.71&                   \\
     305523&O9.0 II    &ASCC 65          & 7.09 &3532& 7.5& 6.3& 27.8&-5.3 [-5.7, -5.1]&0.81&                   \\
     305532&O6.0 V     &ASCC 65          & 7.09 &3448& 3.6& 6.7& 58.0&-4.3 [-4.6, -4.0]&0.80&                   \\
     305539&O7.0 V     &Trumpler 14      & 6.67 &2752& 1.4& 1.3& 21.3&-4.1 [-4.2, -3.6]&0.83&                   \\
\hline\hline
\end{longtable}
Comments:\\
(1) - runaway O-stars according to \citet{2007astro.ph..3005S}\\
(2) - member of the ASCC 45 cluster according to \citet{2005A&A...440..403K}\\
(3) - member of the ASCC 79 cluster according to \citet{2005A&A...440..403K}\\
\end{document}